# Verified residual-specific explicit derivative kernels for physics-informed learning and discretized PDE adjoints

Wenbo Cao[a], Zhe Lu[b,c,d], Weiwei Zhang[b,c,d,]*

[a] *Institute of AI for Industries, Chinese Academy of Sciences, Nanjing 211135, China*

[b] *School of Aeronautics, Northwestern Polytechnical University, Xi'an 710072, China*

[c] *International Joint Institute of Artificial Intelligence on Fluid Mechanics, Northwestern Polytechnical University, Xi'an, 710072, China*

[d] *National Key Laboratory of Aircraft Configuration Design, Xi'an 710072, China*

* *Corresponding author. E-mail address: aeroelastic@nwpu.edu.cn*

**Abstract.** Derivative computation is a central component of scientific computing, ranging from space-time derivatives in physics-informed neural networks (PINNs) to residual Jacobian actions and discrete-adjoint operators in computational fluid dynamics (CFD) solvers. General-purpose automatic differentiation (AD) greatly reduces implementation effort, but it can incur substantial runtime and memory overhead for high-order residuals and complex discretized operators. Explicit derivative kernels can directly exploit problem-specific structure, making derivative evaluations efficient, controllable, and structure-preserving, but they have long been limited by derivation, implementation, and maintenance costs. This work revisits explicit differentiation (ED) as a residual-specific and verifiable derivative-evaluation route enabled by agent-assisted implementation and stringent numerical verification. For PINNs, we propose a residual-specific partial-jet derivative propagation strategy that makes the derivative-state closure of the target PDE residual explicit and realizes it through specialized layerwise kernels, rather than relying solely on nested AD or a generic Taylor-mode transform. Relative to nested AD, the resulting ED kernels achieve floating-point-level agreement in residual and parameter-gradient evaluations, and provide problem-dependent acceleration of complete PINN training, often reaching 2-4× speedups while also reducing peak GPU memory in most cases. As a second realization for discretized PDE adjoints, we apply the same verification-driven explicit-kernel philosophy to a finite-volume CFD residual, where generated tangent-action and transpose-action kernels pass Taylor-remainder, inner-product, and reduced-gradient consistency checks and are embedded into a GPU-resident discrete-adjoint workflow for freestream Mach-number and angle-of-attack inversion. These results suggest that verified explicit derivative kernels, supported by agent-assisted implementation, can provide a practical and structure-aware complement to general-purpose AD for derivative-intensive scientific computing.

## 1 Introduction

Derivative computation is one of the most fundamental algorithmic components in scientific computing. Whether one evaluates space-time derivatives in physics-informed neural networks [1], or implements residual Jacobians together with their tangent-linear and discrete-adjoint actions in conventional numerical PDE solvers, derivative computation directly affects the efficiency, stability, and scalability of the resulting numerical methods. In implicit time marching [2], Newton-Krylov methods [3], sensitivity analysis [4], resolvent analysis [5], and design optimization [6, 7], a high-quality derivative implementation determines not only the cost of a single operator evaluation, but also the accuracy of the linearization, the consistency of the computed gradients, and the convergence behavior of the overall iterative process. Therefore, how to implement derivatives in an efficient, reliable, and maintainable manner has long been a central issue in scientific computing software and algorithm design.

Derivative computation in scientific computing can be implemented through several routes, each with its own trade-offs. Finite-difference approximations are simple and non-intrusive, but they are approximate, sensitive to step-size selection, and become inefficient when many derivative directions are required. Symbolic or fully analytic differentiation can provide explicit formulas, but often becomes cumbersome for large neural networks, complex residual operators, and legacy solver codes. Automatic differentiation [8] offers a more general and systematic alternative, and has been widely adopted in deep learning, differentiable programming, physics-informed neural networks, and PDE-constrained optimization. By automatically applying the chain rule to a computational graph, automatic differentiation substantially lowers the barrier to derivative derivation and code implementation, allowing researchers to rapidly construct differentiable programs involving complex models, loss functions, and physical residuals. This trend is evident in physics-informed learning, where widely used PINN or physics-ML libraries such as DeepXDE [9], SciANN [10], and NVIDIA SimNet/Modulus [11] rely on AD to construct equation residuals and train physics-constrained neural models. It is also evident in aerodynamic design optimization, where adjoint-oriented CFD frameworks such as SU2 [12], ADflow [13], and DAFoam [14] use algorithmic differentiation, source-transformation tools, or structured discrete-adjoint implementations to obtain residual, objective-function, and sensitivity information required for discrete-adjoint analysis and optimization.

However, the convenience of general-purpose automatic differentiation does not imply that it is always the most efficient or suitable implementation strategy in scientific computing. In physics-informed neural networks, high-order space-time derivatives often require nested automatic differentiation calls, which can generate complex computational graphs and lead to significant time and memory overhead [15-17]. In conventional PDE solvers, residual Jacobians and their tangent-linear and discrete-adjoint actions usually possess clear sparsity patterns and local coupling structures. Directly relying on general-purpose automatic differentiation may fail to fully exploit these structures, and integrating AD-generated derivative code into existing high-performance Fortran, C++, or GPU-resident solver kernels can be nontrivial [18-21].

Compared with general-purpose automatic differentiation, explicit derivative implementations can often exploit the mathematical structure of the problem and the structure of the underlying discretization more directly. In the PINN setting, this motivates the residual-specific partial-jet derivative propagation strategy developed in this work, where the derivative-state closure required by the target PDE residual is made explicit and implemented through layerwise derivative-propagation kernels. This organization avoids repeated nested automatic-differentiation calls and reduces reliance on generic high-order differentiation transforms, while enabling residual-guided kernel specialization and fused batched linear operations. In finite-volume or finite-element solvers, residual Jacobian blocks, Jacobian-vector products, and transpose-Jacobian-vector products can be accumulated according to local stencil or element templates, and can be naturally coupled with existing sparse-matrix structures, preconditioners, and adjoint-solution procedures [6, 18]. As a result, explicit derivative kernels and hand-crafted tangent/adjoint implementations can retain important advantages in computational efficiency, memory control, structure preservation, and integration with production-level solvers. However, this route has long been limited by the complexity of derivative derivation, the length and fragility of the resulting code, the risk of implementation errors, and the difficulty of long-term maintenance. As neural-network models, governing equations, and discretization schemes become increasingly complex, implementing explicit derivative kernels by hand often requires substantial domain expertise and development effort, making this approach difficult to use broadly in rapidly evolving scientific computing research.

Recent advances in programming agents and large language models offer a new opportunity to revisit this trade-off [22, 23]. Unlike conventional hand derivation and hand coding, an agent-assisted workflow can generate derivative-propagation code, Jacobian-action code, and adjoint-accumulation code from a prescribed network architecture, residual expression, or local discretization operator, and can further revise the implementation based on feedback from failed tests, following the broader test- and feedback-driven refinement paradigm explored in recent coding-agent studies [24-26]. More importantly, this generation process can be coupled with stringent numerical verification procedures, including finite-difference checks, Taylor-remainder tests, adjoint inner-product consistency tests, and local comparisons with automatic-differentiation results. In this way, explicit derivative kernels no longer need to rely solely on one-shot manual derivation; instead, their implementation can be organized as an agent-assisted, verification-driven generate–verify–revise loop.

Therefore, this work does not aim to propose a new automatic code-generation algorithm or a general-purpose differentiation system. Instead, it revisits the practical value of explicit derivative kernels as verifiable, problem-structured computational modules under agent-assisted implementation and stringent numerical verification. The contributions are twofold. First, we formulate a residual-specific partial-jet derivative propagation strategy for PINN residual construction and evaluate it in complete PINN training under matched automatic-differentiation (AD) and explicit-differentiation (ED) settings. Second, we construct a finite-volume discrete-adjoint realization by developing explicit tangent-action and transpose-action kernels for CFD residuals and embedding them into a GPU-resident workflow for freestream Mach-number and angle-of-attack inversion. Across both settings, programming-agent assistance is used to reduce the implementation and revision burden of explicit derivative kernels, while correctness is established by independent numerical verification rather than by the agent output itself. Together, these two settings show that explicit derivative kernels can become practical scientific-computing modules when residual-specific formulation is combined with agent-assisted implementation and independent verification.

# 2 Methods

## 2.1 Explicit derivative kernels in scientific computing

This work focuses on derivative kernels in scientific computing that possess clear mathematical structure and local coupling patterns. We use the term explicit derivative kernel to denote a specialized computational routine that directly evaluates prescribed derivative quantities for a given neural network or discretized PDE operator. Examples include derivatives of a neural-network output with respect to input space–time coordinates, residual Jacobian actions with respect to discrete state variables, and the corresponding transpose-Jacobian actions. These kernels usually do not require the formation of a full global Jacobian matrix. Instead, they are embedded into residual evaluation, implicit solution procedures, sensitivity analysis, and adjoint-based optimization in the form of local derivative propagation, Jacobian-vector products, or transpose-Jacobian-vector products.

Compared with general-purpose automatic differentiation, the purpose of explicit derivative kernels is not to provide black-box differentiation for arbitrary computational graphs. Rather, it is to exploit structural information that is specific to a given scientific computing problem. For physics-informed neural networks, once the network architecture and activation function are fixed, the types of space–time derivatives required by the governing residual are usually known in advance. These derivatives can therefore be computed directly through layerwise chain-rule propagation. For CFD residual operators, the coupling between control volumes is restricted to neighboring cells and boundary faces, so residual Jacobian blocks and their transpose actions can be accumulated according to local flux, boundary, and viscous templates. Based on this observation, this work examines whether explicit derivative kernels, when generated with agent assistance and subjected to stringent numerical verification, can become a practical and computationally efficient route for derivative implementation in scientific computing.

**2.2 Residual-specific partial-jet propagation for PINN residuals**

In physics-informed neural networks, PDE residuals depend on derivatives of the neural-network output with respect to spatial and temporal coordinates. In this work, we formulate a residual-specific partial-jet derivative propagation strategy for constructing these residuals. The central idea is to make the derivative-state closure required by the target PDE residual explicit and to realize this closure through layerwise propagation kernels, rather than repeatedly differentiating the completed network output through nested AD or relying on a generic high-order differentiation transform. Let the neural network take the coordinate vector $\boldsymbol{x} = (x_1, \ldots, x_d)$ as input and output the physical state $u_\theta(\boldsymbol{x})$. For a given PDE operator $\mathcal{N}(\cdot)$, the residual can be written as $r_\theta(\boldsymbol{x}) = \mathcal{N}\left(u_\theta, \partial_{x_i} u_\theta, \partial_{x_i x_j} u_\theta, \ldots\right)$. Conventional PINN implementations usually compute these coordinate derivatives through automatic differentiation. This approach is convenient and general, but it can become expensive when the residual contains high-order derivatives, multiple output fields, or a large number of collocation points, because nested differentiation may create deep computational graphs and substantial memory overhead. In this work, we refer to the conventional implementation based on nested automatic differentiation as the AD implementation, and to the implementation based on explicit layerwise coordinate-derivative propagation as explicit differentiation (ED). Here, ED denotes only the explicit evaluation of coordinate derivatives entering the PDE residual, not a replacement of reverse-mode parameter differentiation used in network training.

For feedforward neural networks, the required coordinate derivatives can instead be evaluated by propagating derivative states through the network together with the forward pass. This avoids repeatedly differentiating the completed network output and turns coordinate differentiation into a layerwise chain-rule propagation problem. Consider the $\ell$-th layer,

$$\boldsymbol{z}^\ell = \boldsymbol{W}^\ell \boldsymbol{a}^{\ell-1} + \boldsymbol{b}^\ell, \boldsymbol{a}^\ell = \phi(\boldsymbol{z}^\ell) \tag{1}$$

where $\phi$ is an elementwise activation function. If $\boldsymbol{a}^{\ell-1}$ and its required derivatives with respect to the input coordinates are available from the previous layer, the corresponding derivative states at the current layer can be obtained directly from the chain rule. For example, the first-order derivatives satisfy

$$\partial_{x_i} \boldsymbol{z}^\ell = \boldsymbol{W}^\ell \partial_{x_i} \boldsymbol{a}^{\ell-1}, \partial_{x_i} \boldsymbol{a}^\ell = \phi'(\boldsymbol{z}^\ell) \odot \partial_{x_i} \boldsymbol{z}^\ell \tag{2}$$

where $\odot$ denotes elementwise multiplication. The second-order derivatives are propagated as

$$\partial_{x_i x_j} \boldsymbol{z}^\ell = \boldsymbol{W}^\ell \partial_{x_i x_j} \boldsymbol{a}^{\ell-1}, \partial_{x_i x_j} \boldsymbol{a}^\ell = \phi''(\boldsymbol{z}^\ell) \odot \partial_{x_i} \boldsymbol{z}^\ell \odot \partial_{x_j} \boldsymbol{z}^\ell + \phi'(\boldsymbol{z}^\ell) \odot \partial_{x_i x_j} \boldsymbol{z}^\ell \tag{3}$$

Higher-order coordinate derivatives can be constructed recursively in the same manner by extending the set of propagated derivative states. Thus, in the ED implementation, the derivatives required by the PINN residual are obtained during a single explicit derivative-propagation pass, rather than through multiple nested automatic-differentiation calls.

It is important to distinguish this procedure from a global symbolic expansion of the entire neural-network expression. The network is still evaluated layer by layer, and each layer is equipped with a local derivative-propagation rule that follows its computational structure. This makes the implementation naturally compatible with vectorized evaluation and batched collocation points. In practice, we adopt an on-demand partial-jet strategy: only the derivative states required by the target PDE residual are propagated, while the full Hessian or irrelevant mixed derivatives are not constructed. For example, a second-order residual may only require the states $(u, u_t, u_x, u_{xx})$, whereas a fourth-order residual further requires the lower-order derivative chain needed to obtain $u_{xxxx}$. In the linear layers, the same weight matrix acts on the function value and on all propagated derivative states. We therefore concatenate the required states along the batch dimension, apply a fused GEMM operation for the linear propagation, and then split the results before applying the activation-layer chain-rule updates. This implementation preserves the exact chain-rule relation for the prescribed coordinate derivatives while organizing the required derivative states explicitly, avoiding repeated coordinate-AD calls, and reducing small matrix-multiplication operations through batched linear propagation.

The ED implementation used here is closely related to forward-mode and Taylor-mode automatic differentiation. In Taylor-mode AD, truncated Taylor polynomial coefficients are propagated through the computational graph, allowing high-order derivatives to be obtained without repeatedly applying first-order AD and thereby reducing the redundant work that may arise in nested high-order differentiation [16, 17]. This advantage is most relevant when high-order derivatives are required; for purely first-order derivative evaluations, the relative benefit is more problem- and implementation-dependent. The ED kernels in this work follow the same derivative-state propagation principle, but they are not intended as a general-purpose Taylor-mode transform. Instead, the set of propagated states is selected from the structure of the target PDE residual and then implemented as an explicit residual-specific kernel, as described above. Thus, ED should be understood as a problem-specific partial-jet realization of forward derivative propagation, rather than as a new differentiation calculus or a replacement for general-purpose AD.

**2.3 Explicit derivative kernels for discrete PDE residuals**

Beyond coordinate derivatives in physics-informed neural networks, another important class of explicit derivative kernels arises from discrete PDE residual operators. These kernels provide the tangent and transpose derivative actions required by implicit linearization, sensitivity analysis, and discrete-adjoint optimization. Let $\boldsymbol{q}$ denote the discrete state variables and $\boldsymbol{\mu}$ denote a set of design or control variables. A steady discrete PDE system can be written as

$$\boldsymbol{R}(\boldsymbol{q}, \boldsymbol{\mu}) = \boldsymbol{0} \tag{4}$$

In implicit solution methods, sensitivity analysis, and PDE-constrained optimization, one typically needs derivative actions of the residual with respect to both the state variables and the parameters, such as

$$R_q \boldsymbol{v}, R_\mu \boldsymbol{s}, R_q^T \boldsymbol{w}, R_\mu^T \boldsymbol{w} \tag{5}$$

where

$$R_q = \frac{\partial \boldsymbol{R}}{\partial \boldsymbol{q}}, R_\mu = \frac{\partial \boldsymbol{R}}{\partial \boldsymbol{\mu}} \tag{6}$$

Here, $\boldsymbol{v}$ and $\boldsymbol{s}$ denote perturbations in the state and parameter spaces, respectively, and $\boldsymbol{w}$ is a vector in the residual space used for transpose-Jacobian actions. These derivative actions do not necessarily require the assembly and storage of a full global Jacobian matrix. In finite-volume, finite-element, or related discretizations, the residual is usually assembled from local contributions associated with cells, faces, boundaries, or elements. The same locality can be exploited to construct explicit derivative kernels for tangent and transpose actions.

In the CFD residual considered in this work, the residual is evaluated by looping over internal faces and boundary contributions. For an internal face, the numerical flux depends on the local states on the two sides of the face and, when relevant, physical or freestream parameters. If a face contribution is written abstractly as

$$\boldsymbol{F}_f = \boldsymbol{F}_f(\boldsymbol{q}_L, \boldsymbol{q}_R, \boldsymbol{\mu}) \tag{7}$$

then the corresponding local tangent action takes the form

$$\delta \boldsymbol{F}_f = F_{f,L} \delta \boldsymbol{q}_L + F_{f,R} \delta \boldsymbol{q}_R + F_{f,\mu} \delta \boldsymbol{\mu} \tag{8}$$

where $F_{f,L}, F_{f,R}$, and $F_{f,\mu}$ denote local derivative blocks. These local tangent contributions are accumulated into the residual perturbations of the adjacent control volumes, yielding the global actions $R_q \boldsymbol{v}$ and $R_\mu \boldsymbol{s}$. The transpose kernel performs the corresponding reverse accumulation: a residual-space vector is distributed back to the local state and parameter variables through the transposed local derivative blocks $F_{f,L}^T, F_{f,R}^T$, and $F_{f,\mu}^T$. Boundary-face terms are handled in the same manner, with local templates determined by the corresponding boundary condition. Thus, the global transpose actions $R_q^T \boldsymbol{w}$ and $R_\mu^T \boldsymbol{w}$ are evaluated by accumulating local transpose contributions over the same residual components used by the primal discretization.

The implementation follows this local derivative-kernel view. For each residual component, the generated kernel provides either a tangent action, a transpose action, or both, depending on its role in linearization and adjoint computation. Local derivative blocks may be constructed on the fly for verification or local operator application, but the full global residual Jacobian is not assembled as a stored sparse matrix. Instead, the generated kernels provide matrix-free evaluations of the global residual Jacobian actions and transpose-Jacobian actions. This design allows the derivative implementation to remain consistent with the primal residual discretization while still being compatible with existing residual evaluation, linearization, and adjoint-solution workflows.

For a scalar objective function $J(\boldsymbol{q}, \boldsymbol{\mu})$, the discrete adjoint formulation provides an efficient way to compute design variable gradients when the state variables satisfy $\boldsymbol{R}(\boldsymbol{q}, \boldsymbol{\mu}) = \boldsymbol{0}$. Introducing the adjoint variable, or Lagrange multiplier, $\boldsymbol{\lambda}$, and defining the Lagrangian as

$$\mathcal{L}(\boldsymbol{q}, \boldsymbol{\mu}, \boldsymbol{\lambda}) = J(\boldsymbol{q}, \boldsymbol{\mu}) + \boldsymbol{\lambda}^T \mathbf{R}(\boldsymbol{q}, \boldsymbol{\mu}) \tag{9}$$

the discrete adjoint equation is

$$R_q^T \boldsymbol{\lambda} = -J_q^T \tag{10}$$

and the corresponding total derivative with respect to the parameters is

$$\frac{dJ}{d\boldsymbol{\mu}} = J_\mu + \boldsymbol{\lambda}^T R_\mu \tag{11}$$

Therefore, the central computational requirement in adjoint-based optimization is not the explicit storage of $R_q$ or $R_\mu$ as global matrices, but the reliable evaluation of $R_q^T \boldsymbol{\lambda}$, $R_\mu^T \boldsymbol{\lambda}$, and the associated objective derivatives. The role of explicit derivative kernels is to provide these tangent and transpose derivative actions in a form that is consistent with the face-based primal residual and directly embeddable into existing linearization or adjoint workflows.

**2.4 Verification of explicit derivative kernels**

Because explicit derivative kernels may be generated with agent assistance, their correctness cannot be assumed from the code structure or from manual inspection alone. In this work, each generated kernel is subjected to independent numerical consistency tests before it is used in the subsequent experiments. The verification procedures are chosen according to the type of derivative being evaluated: coordinate derivatives in neural networks, tangent actions of discrete residuals, transpose-Jacobian actions, and adjoint-based parameter gradients.

For the coordinate-derivative kernels used in physics-informed neural networks, we first compare the ED results with those obtained from AD. The comparison is performed at randomly sampled input points and includes the network output and the coordinate derivatives required by the target PDE residual, such as $u, u_t, u_x, u_{xx}$, and higher-order derivatives when needed. The relative errors between the ED and AD are used to confirm that the layerwise propagation rules, activation derivatives, and residual assembly are consistent with the chain rule.

For discrete PDE residuals, the tangent kernels are verified against the original nonlinear residual through finite-difference directional-derivative checks and Taylor-remainder tests. Given a state perturbation $\boldsymbol{v}$, the finite-difference directional derivative

$$\frac{\boldsymbol{R}(\boldsymbol{q}+\varepsilon\boldsymbol{v},\boldsymbol{\mu})-\boldsymbol{R}(\boldsymbol{q},\boldsymbol{\mu})}{\varepsilon} \tag{12}$$

is compared with the explicit tangent-kernel output $R_q\boldsymbol{v}$ over a range of perturbation sizes $\varepsilon$. Equivalently, the Taylor expansion

$$\boldsymbol{R}(\boldsymbol{q}+\varepsilon\boldsymbol{v},\boldsymbol{\mu})=\boldsymbol{R}(\boldsymbol{q},\boldsymbol{\mu})+\varepsilon R_q\boldsymbol{v}+O(\varepsilon^2) \tag{13}$$

implies that the remainder

$$E_q(\varepsilon)=\left\|\boldsymbol{R}(\boldsymbol{q}+\varepsilon\boldsymbol{v},\boldsymbol{\mu})-\boldsymbol{R}(\boldsymbol{q},\boldsymbol{\mu})-\varepsilon R_q\boldsymbol{v}\right\| \tag{14}$$

should decrease quadratically with $\varepsilon$ before round-off errors dominate. The same procedure is applied to parameter-derivative kernels by perturbing $\boldsymbol{\mu}$ along a direction $\boldsymbol{s}$ and comparing the finite-difference directional derivative with $R_\mu\boldsymbol{s}$. The transpose-Jacobian kernels are verified by adjoint inner-product consistency tests. For arbitrary perturbations $\boldsymbol{v}$ and $\boldsymbol{s}$, and for an arbitrary residual-space vector $\boldsymbol{w}$, the following identities should hold

$$\left\langle R_q\boldsymbol{v},\boldsymbol{w}\right\rangle=\left\langle \boldsymbol{v},R_q^T\boldsymbol{w}\right\rangle,\left\langle R_\mu\boldsymbol{s},\boldsymbol{w}\right\rangle=\left\langle \boldsymbol{s},R_\mu^T\boldsymbol{w}\right\rangle \tag{15}$$

These tests directly check whether the transpose kernels are consistent with the corresponding tangent kernels under the same discrete inner product. They are especially important for adjoint-based computations, because a tangent kernel that passes a Taylor test does not automatically guarantee that the implemented transpose accumulation is correct.

### 2.5 Implementation notes on agent assistance

The agent-assisted component in this work should be understood as an implementation workflow rather than a separate algorithmic contribution. In practice, the programming agent was provided with the existing primal code, the target derivative task, and the corresponding verification procedure. For the PINN examples, the target tasks were the explicit coordinate-derivative propagation kernels and the associated residual evaluations in PyTorch/JAX. For the CFD example, the target tasks were the matrix-free tangent-action and transpose-accumulation kernels consistent with the existing Fortran-based GPU-resident finite-volume residual implementation. The agent was used to generate or revise candidate implementations, while runtime errors, tensor-shape inconsistencies, AD-comparison errors, Taylor-remainder behavior, transpose inner-product errors, and reduced-gradient discrepancies were used as feedback during revision. The final kernels used in the experiments are the accepted implementations that pass the verification procedures described above. Thus, the agent-assisted workflow reduces the practical implementation and revision cost of explicit derivative kernels, whereas correctness is established by independent numerical verification rather than by the agent output itself.

## 3 Numerical experiments

This section evaluates the verified explicit derivative kernels in two representative realizations. The first setting concerns coordinate-derivative evaluation in physics-informed neural networks, where we compare ED with AD for high-order PDE residuals. The second setting concerns derivative actions of discrete CFD residuals, where the generated tangent and transpose kernels are verified and then embedded into a GPU-resident discrete-adjoint workflow for freestream-parameter optimization.

All generated kernels are subjected to the verification procedures described in Section 2.4 before being used in the numerical experiments. The PINN examples focus on derivative accuracy and end-to-end training efficiency, whereas the CFD example focuses on the consistency of residual derivative kernels and their integration into adjoint-based Mach-number and angle-of-attack optimization.

### 3.1 PINN training with residual-specific partial-jet propagation

#### 3.1.1 Problem setup and matched baselines

We evaluate ED in complete PINN training on three benchmark residuals with different derivative structures. The first case is the one-dimensional viscous Burgers equation on $(x,t)\in[0,1]\times[0,1]$, $r=u_t+uu_x-0.01u_{xx}$. The loss consists of a weighted initial-condition term, periodic boundary terms for both $u$ and $u_x$, and the PDE residual term. We use 101 initial-condition points, 1000 periodic boundary time samples, and 10000 PDE collocation points. The second case is the two-dimensional lid-driven incompressible Navier-Stokes problem on the unit square at $Re$ = 100. The network outputs $(u,v,p)$, and the residuals are

$$\begin{aligned}
r_c &= u_x+v_y \\
r_u &= uu_x+vu_y+p_x-(u_{xx}+u_{yy})/Re \\
r_v &= uv_x+vv_y+p_y-(v_{xx}+v_{yy})/Re
\end{aligned} \tag{16}$$

The loss combines velocity boundary conditions on the lid and no-slip walls with the three PDE residuals, using 10000 PDE collocation points, 1499 wall points, and 499 lid points. The third case is the Kuramoto-Sivashinsky (KS) equation, $r=u_t+5uu_x+0.5u_{xx}+0.005u_{xxxx}$. We train on the first time window of the reference solution, using 512 initial-condition points and 4096 PDE collocation points. Periodicity in the KS case is imposed through the input embedding rather than by a separate boundary loss.

For the Burgers and Navier-Stokes cases, we use tanh multilayer perceptrons with six hidden layers and 64 neurons per hidden layer. For the KS equation, we use a modified MLP with periodic/Fourier input embedding, following the periodic/Fourier-embedding setting used in previous PINN training studies, because a standard PINN architecture is difficult to train stably for this problem [27]. The AD and ED implementations are compared under matched conditions: they use the same network architecture, initial parameters, collocation points, loss definition, dtype, device, optimizer, and number of training steps. The only difference is how the coordinate derivatives entering the PDE residual are evaluated. The AD baseline obtains these derivatives through nested automatic differentiation, whereas ED propagates the required coordinate-derivative states explicitly through the network layers. In all cases, parameter gradients for training are still computed by reverse-mode differentiation of the scalar loss; ED replaces only the coordinate-derivative evaluation inside the PDE residual.

In all three PINN examples, we use an all-points batching layout: the initial-condition, boundary-condition, and PDE collocation points are concatenated before network evaluation and coordinate-derivative computation, and the outputs are split afterward to assemble the loss terms. This layout is used for both AD and ED. Although it may compute some derivatives at points where they are not directly used, it avoids multiple small forward/derivative evaluations and improves GPU throughput.

Unless otherwise stated, all PINN timing experiments are performed on a single NVIDIA GeForce RTX 5080 GPU using float32 and full-batch Adam with a learning rate of $10^{-3}$. Each timing result is averaged over three independent runs after 20 warm-up steps, with GPU synchronization applied before and after timing. The measured runtime includes residual construction, scalar loss evaluation, reverse-mode parameter-gradient computation, and optimizer updates, thereby reflecting the practical end-to-end cost observed by a PINN user rather than the residual-forward cost alone. Runtime and memory comparisons are made within the same backend under matched precision, device, network, data, loss, and optimizer settings.

### 3.1.2 Verification, training trajectories, and computational cost

Before comparing runtime and memory usage, we first verify whether ED reproduces the AD-based residual evaluation. We directly compare the residual vectors and the parameter gradients obtained by differentiating the scalar residual loss. As summarized in Table 1, the two implementations agree to the corresponding floating-point accuracy: in single precision, the residual and gradient discrepancies remain at the $10^{-8}$-$10^{-6}$ level, with gradient cosine similarities essentially equal to one; in the double-precision Burgers diagnostic, the discrepancies further decrease to near machine precision. These results confirm that ED reproduces the AD residuals and the resulting training gradients up to finite-precision round-off.

Table 1. Consistency between AD- and ED-based PINN residual evaluations.

| Problem | dtype | Residual rel. $L_2$ error | Gradient rel. $L_2$ error | Gradient cosine |
|---|---|---|---|---|
| Burgers | float32 | $2.18\times10^{-8}$ | $6.08\times10^{-7}$ | 0.99999994 |
| KS | float32 | $7.72\times10^{-7}$ | $3.28\times10^{-7}$ | 1.00000000 |
| NS | float32 | $4.95\times10^{-7}$ | $2.60\times10^{-7}$ | 1.00000000 |
| Burgers | float64 | $4.28\times10^{-17}$ | $7.77\times10^{-15}$ | 1.00000000 |

We next examine the actual PINN optimization process. Figure 1 compares the 1000-step Adam loss histories obtained with AD- and ED-based residual evaluations. For the KS case, the two curves are visually indistinguishable in single precision. The Burgers and NS cases show noticeable differences in the intermittent loss spikes under float32, although the overall decay trend remains consistent. When the same Burgers test is repeated in float64, the AD and ED trajectories nearly coincide. This indicates that the float32 discrepancy is caused by finite-precision sensitivity of the Adam trajectory, rather than by an inconsistency in the explicit derivative formulas. Thus, the training histories provide an end-to-end check that is consistent with the residual- and gradient-level verification in Table 1.

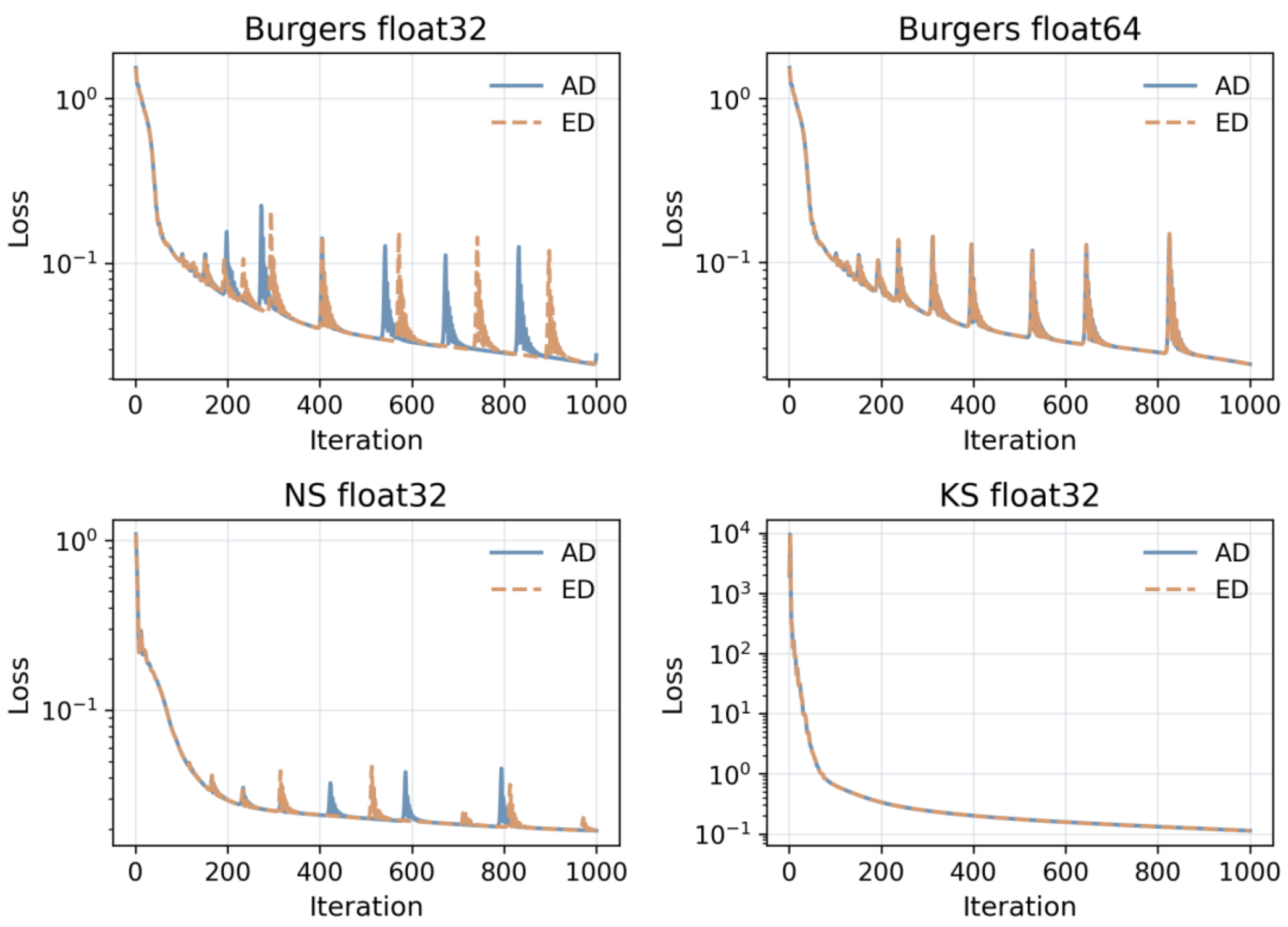

Figure 1. Adam loss histories obtained using AD- and ED-based PINN residual evaluations.

Figure 2 reports the cost of complete 1000-step full-batch Adam training, including residual evaluation, scalar loss construction, reverse-mode parameter-gradient computation, and optimizer updates. ED reduces the total training time for all three problems, with speedups of 1.22×, 2.30×, and 3.58× for Burgers, NS, and KS, respectively. The largest gain appears in the KS case, where the fourth-order derivative makes nested coordinate AD especially costly. The peak GPU memory shows a problem-dependent behavior: it is substantially reduced for the NS and KS cases, whereas a slight increase is observed for the simpler Burgers residual. This behavior indicates that the memory benefit of ED depends on the balance between the graph-storage overhead avoided from nested coordinate AD and the additional intermediate states introduced by explicit derivative propagation. Overall, these results show that replacing coordinate AD with ED inside the residual can reduce the practical PINN training cost, especially for residuals with more expensive high-order or multi-field derivative structures, without changing the loss function, optimizer, or reverse-mode parameter-gradient mechanism.

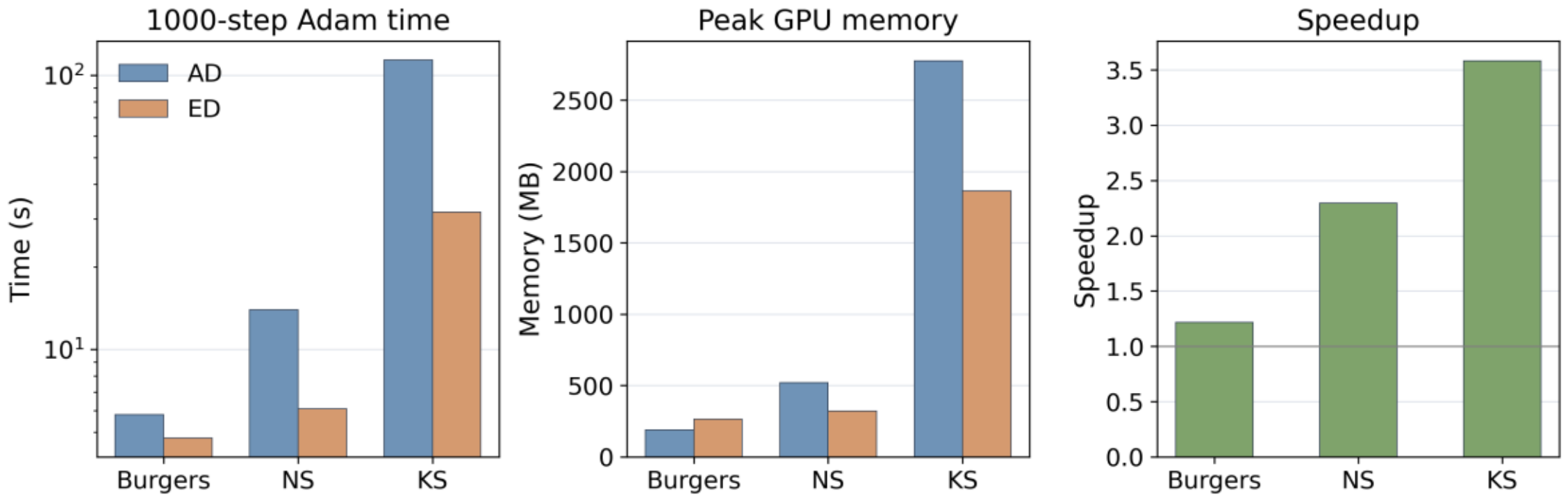

Figure 2. PyTorch runtime, peak GPU memory, and speedup for AD- and ED-based full PINN training. The runtime corresponds to 1000 full-batch Adam steps.

Figure 3 further examines the sensitivity of the end-to-end training speedup to the number of PDE collocation points and the hidden-layer width. In these tests, the reported time again corresponds to complete 1000-step full-batch Adam training. When the number of PDE collocation points is varied from 1024 to 16384, the speedup remains nearly unchanged for the Burgers and NS cases, staying around 1.2× and 2.2–2.3×, respectively. The KS case consistently shows the largest acceleration, around 3.8–4.2×, because its fourth-order derivative makes nested coordinate AD particularly expensive. When the hidden width is varied from 16 to 128, the Burgers and NS cases remain relatively stable, whereas the KS case exhibits a larger variation, with the largest speedup observed at width 64. This variation is expected because both the nested-AD baseline and the ED implementation for the KS residual involve more complex high-order derivative graphs or derivative-state propagation, making the measured speedup more sensitive to tensor shapes, GPU kernel scheduling, and memory behavior. Overall, these results indicate that the ED acceleration is robust with respect to collocation density and network width, while the absolute speedup depends on the derivative complexity of the target residual, backend-level transformation overhead, tensor shapes, and GPU kernel scheduling.

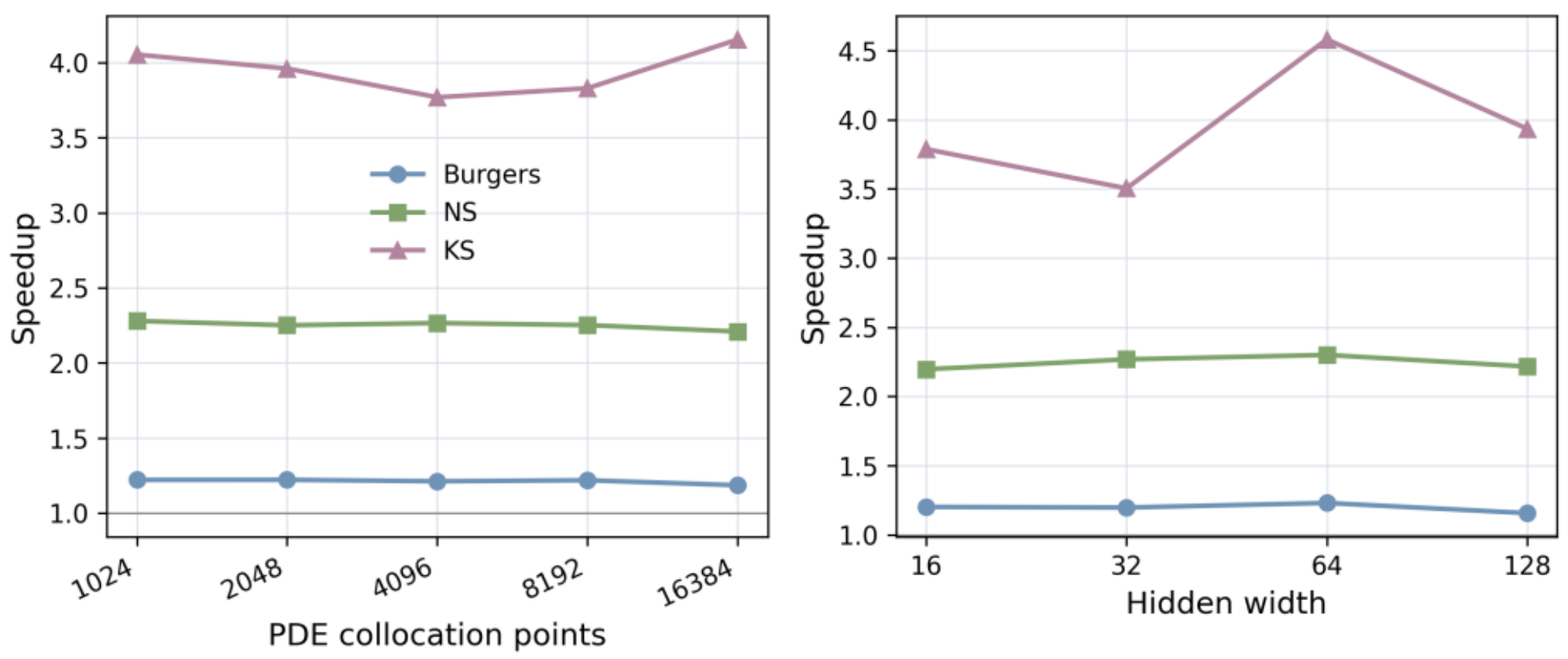

Figure 3. Sensitivity of end-to-end PINN training speedup to collocation points and hidden width.

### 3.1.3 Comparison with Taylor-mode AD on a compiled backend

To further distinguish ED from general Taylor-mode AD, we repeat the complete 1000-step full-batch Adam training benchmark in JAX with XLA compilation and include a Taylor-mode AD baseline. The Taylor-mode AD baseline is implemented using coordinate-direction Taylor jets: temporal and spatial coordinate directions are seeded to obtain the derivatives required by each PDE residual, while the network, collocation points, loss function, optimizer, and training protocol are kept the same as in the nested-AD and ED cases.

As shown in Figure 4, Taylor-mode AD already reduces the runtime relative to nested AD, with speedups of 1.31×, 1.47×, and 1.76× for Burgers, NS, and KS, respectively. This confirms that propagating higher-order derivative information in a forward Taylor form can reduce part of the overhead caused by repeated nested differentiation. However, ED remains faster than both AD baselines, with speedups of 2.28×, 2.27×, and 3.01× for Burgers, NS, and KS, respectively. ED also gives the lowest peak GPU memory in all three JAX cases. The absolute wall-clock times in JAX are substantially lower than those in the PyTorch experiments because XLA compilation can fuse operations and optimize the residual-evaluation graph. Therefore, the PyTorch and JAX results should not be interpreted as a direct backend comparison; the relevant comparison is made within each backend. Within the compiled JAX backend, the results show that the advantage of ED is not merely a consequence of replacing nested reverse-mode AD with a forward high-order AD transform. Instead, the measured gain reflects the combined effect of explicit layerwise derivative propagation, residual-guided kernel specialization, reduced framework-level differentiation overhead, and fused batched linear operations.

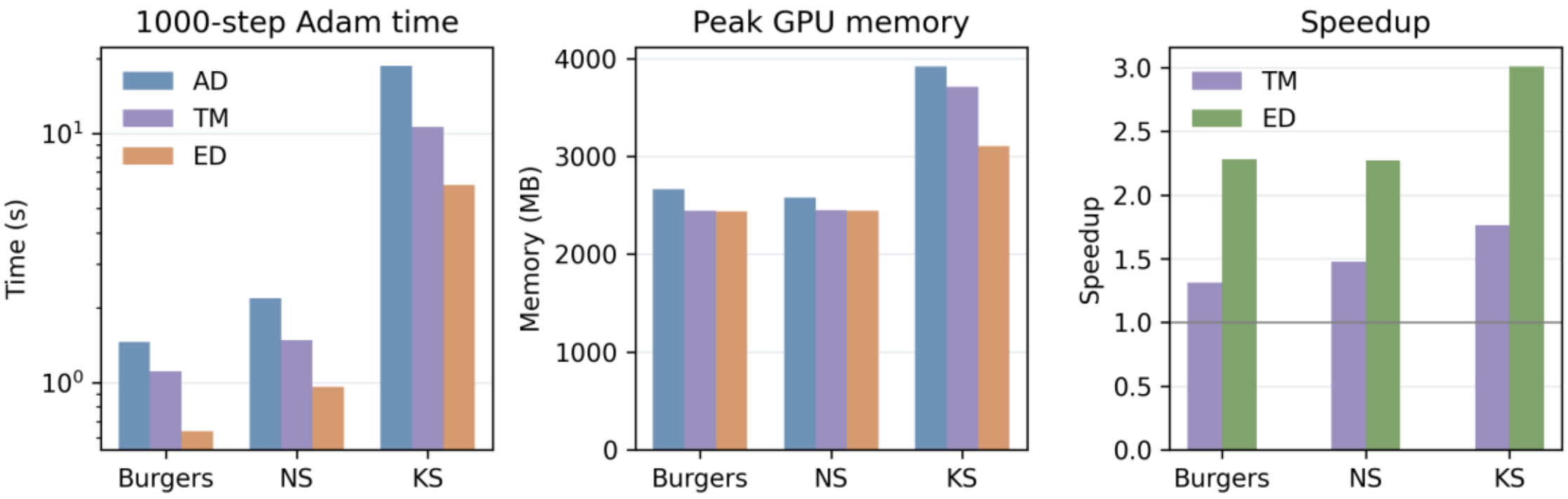

Figure 4. JAX/XLA comparison of nested AD, Taylor-mode AD, and ED for full PINN training. TM denotes Taylor-mode AD implemented with coordinate-direction Taylor jets. Speedup is measured relative to the nested-AD baseline in the same JAX/XLA backend.

To interpret the performance differences in Figure 4, Table 2 compares how the three approaches construct coordinate derivatives inside the PINN residual. Taylor-mode AD reduces the overhead of nested AD by propagating Taylor jets with prescribed orders and coordinate seeds, whereas ED further uses the known PDE residual to construct an explicit residual-specific derivative closure.

Table 2. Comparison of derivative-evaluation strategies for PINN residuals. Note: “AD-transform overhead” refers to the use of general AD transformations for coordinate-derivative evaluation inside the PDE residual. Parameter gradients for training are still computed by reverse-mode AD in all cases.

| Method | Derivative construction | Residual-aware closure | Graph/transform overhead | Kernel specialization |
|---|---|---|---|---|
| Nested AD | Repeated AD calls | Partial | High | No |
| Taylor-mode AD | Taylor jets | Partial | Moderate | No |
| Proposed ED | Explicit closure | Yes | Low | Yes |

This distinction explains why Taylor-mode AD improves over nested AD, while ED can still provide additional acceleration by realizing the residual derivative structure as an explicit layerwise propagation and residual-assembly kernel rather than as a generic differentiation transform.

## 3.2 CFD freestream-parameter inversion with verified adjoint derivatives

### 3.2.1 Problem setup and adjoint-gradient formulation

The second set of experiments realizes the proposed explicit-derivative-kernel strategy in a conventional finite-volume CFD discrete-adjoint workflow. Unlike the PINN experiments, where the derivative kernels act on neural-network coordinate derivatives, the present example concerns derivative actions of a finite-volume discretization of the steady laminar compressible Navier-Stokes equations with respect to the flow state and freestream parameters. The CFD residual is assembled as a cell-integrated face-flux residual without forming a global Jacobian matrix. Inviscid fluxes are computed using a Roe approximate Riemann solver with entropy correction, and face states are obtained by linear reconstruction based on Green-Gauss gradients. The laminar viscous flux is constructed from velocity and temperature gradients with Sutherland viscosity, and no turbulence model is used in the present test. The main boundary types are no-slip wall boundaries and farfield boundaries, with the farfield state determined by the freestream Mach number and angle of attack. For each design point, the primal steady flow is converged on the GPU using pseudo-time marching before the objective and adjoint equations are evaluated. The generated tangent-action and transpose-action kernels, together with the adjoint-gradient evaluation, are implemented in the same GPU-resident finite-volume framework as the primal residual. This provides a finite-volume realization of the proposed strategy in a high-performance solver environment, rather than a benchmark against AD-generated GPU adjoints.

We consider a freestream-parameter inversion problem over a fixed airfoil geometry, using the wall pressure-coefficient distribution as the observation target. This finite-volume realization is designed to exercise the residual tangent kernels, transpose-action kernels, objective derivatives, and freestream-parameter derivative contributions in a complete discrete-adjoint inversion loop. The goal is not to benchmark CFD optimization speed, which is dominated by repeated primal and adjoint solves, but to demonstrate that the generated explicit derivative kernels can provide consistent adjoint gradients and support parameter optimization in a traditional discretized PDE solver.

The geometry and mesh are kept fixed throughout the optimization. The design variables are the freestream Mach number and angle of attack, $\boldsymbol{\mu} = (M_\infty, \alpha)$, while the Reynolds number is fixed at $Re$ = 1000. The target pressure coefficient distribution is generated by running the primal solver at $M_{\infty,tar} = 0.5, \alpha_{tar} = 3°$ and the inversion is initialized from $M_{\infty,0} = 0.3, \alpha_0 = 0°$. At each design point, the steady flow is solved and the loss function is defined as the mean-squared mismatch between the wall pressure coefficient and the target wall pressure coefficient,

$$J = \frac{1}{N_w} \sum_{i=1}^{N_w} \frac{1}{2} (C_{p,i} - C^*_{p,i})^2 \tag{17}$$

where $N_w$ is the number of wall faces used in the objective, $C_{p,i}$ is the computed pressure coefficient at the $i$-th wall face, and $C^*_{p,i}$ is the corresponding target value. The design-variable gradient is computed using the verified discrete-adjoint derivative kernels described in Section 2.3, and the resulting gradient is supplied to a scaled BFGS optimizer to update $\boldsymbol{\mu}$.

### 3.2.2 Verification of finite-volume derivative kernels

Before performing the parameter inversion, we first verify the generated CFD derivative kernels through a staged consistency procedure. The verification drivers are constructed in the same agent-assisted generate--verify--revise workflow, following the local face-based structure of the original residual implementation. The tangent kernels are checked by comparing the explicit matrix-free $\boldsymbol{Jv}$ action with finite-difference perturbations of the nonlinear residual through Taylor-remainder tests, while the transpose kernels are checked by adjoint inner-product consistency. As summarized in Table 3, the verification proceeds from local flux kernels to globally assembled residual actions, and further to wall-boundary, objective, and farfield parameter-dependent contributions. The Taylor remainders show the expected second-order decay over the listed perturbation ranges before round-off effects dominate, and the transpose inner-product errors remain close to machine precision. These results verify the local derivative formulas, global residual accumulation, reverse transpose accumulation, objective derivative, and design-variable derivative contributions before they are coupled through the discrete-adjoint solve.

Table 3. Staged verification of the generated CFD derivative kernels. The tangent/Taylor check compares the explicit matrix-free $\boldsymbol{Jv}$ action with finite-difference perturbations of the nonlinear residual by evaluating the Taylor remainder. The transpose check verifies the identity $\langle \boldsymbol{Jv}, \boldsymbol{w}\rangle = \langle \boldsymbol{v}, \boldsymbol{J}^{\mathrm{T}}\boldsymbol{w}\rangle$

| Kernel or operator | Observed second-order Taylor range | Transpose inner-product error |
| --- | --- | --- |
| Local Roe flux kernel | until round-off | $1.6\times10^{-15}$ |
| Global internal-face Roe residual | $10^{-1}$-$10^{-4}$ | $7.8\times10^{-14}$ |
| Local laminar viscous flux kernel | until round-off | $7.3\times10^{-14}$ |
| Global laminar viscous residual | $10^{-1}$-$10^{-4}$ | $4.3\times10^{-14}$ |
| Global Roe + viscous residual | $10^{-1}$-$10^{-4}$ | $3.2\times10^{-14}$ |
| Global Roe with wall boundary | $10^{-1}$-$10^{-4}$ | $2.5\times10^{-15}$ |
| Wall $C_p$-mismatch objective | $10^{-1}$-$10^{-4}$ | N/A |

As a final end-to-end validation, we compare the discrete-adjoint reduced gradients with finite-difference reduced gradients at the initial design point. In this test, the finite-difference gradient is obtained by perturbing one design variable at a time, recomputing the steady primal flow, and reevaluating the wall $C_p$ -mismatch loss. Two finite-difference step-size sets are used: $(h_M, h_\alpha) = (0.002, 0.05^\circ)$ and $(0.001, 0.025^\circ)$. As shown in Table 4, the resulting adjoint gradients agree closely with both finite-difference estimates, with relative discrepancies of about 0.17% for $M_\infty$ and 0.23% for $\alpha$. The remaining discrepancies are mainly attributed to finite-difference truncation/round-off effects and the tolerance of the converged primal solutions. This final reduced-gradient check confirms that the verified residual transpose kernels, objective derivatives, and freestream-parameter derivative contributions are consistently integrated into the discrete-adjoint gradient calculation.

Table 4. End-to-end verification of the discrete-adjoint reduced gradients.

| Gradient component | Adjoint | FD, Set 1 | FD, Set 2 | Relative error |
|---|---|---|---|---|
| $dJ / dM_\infty$ | $1.86602\times10^{-2}$ | $1.86300\times10^{-2}$ | $1.86291\times10^{-2}$ | 0.17% |
| $dJ / d\alpha$ | $-1.33951\times10^{-4}$ | $-1.34256\times10^{-4}$ | $-1.34263\times10^{-4}$ | 0.23% |

### 3.2.3 Freestream-parameter inversion

Using the verified adjoint gradients, we perform the freestream-parameter inversion with a scaled BFGS optimizer. Figure 5 shows the resulting optimization history. Starting from $(M_\infty,\alpha)=(0.3,0^\circ)$, the objective decreases from $1.19\times10^{-2}$ to $4.60\times10^{-12}$ after 16 accepted BFGS iterations. The recovered parameters are $M_\infty = 0.499988$ and $\alpha = 2.999942^\circ$, which are very close to the target values $(0.5,3^\circ)$. The convergence history (Figure 5) shows that the objective remains nearly flat during part of the intermediate BFGS iterations and then decreases rapidly once the search direction and inverse-Hessian approximation become well aligned with the reduced objective landscape. The parameter errors exhibit a similar behavior: both the Mach-number and angle-of-attack errors are reduced by several orders of magnitude near the end of the optimization. The wall pressure-coefficient distribution further confirms the quality of the recovered parameters. The initial $C_p$ profile differs noticeably from the target distribution, whereas the optimized profile nearly overlaps with the target over the airfoil surface.

The BFGS inversion completes the finite-volume realization of the proposed explicit-kernel strategy. Together with the Taylor-remainder, transpose inner-product, and reduced-gradient checks reported above, this result shows that the tangent-action kernels, transpose-action kernels, objective derivatives, and freestream-parameter derivative contributions are consistently integrated into the discrete-adjoint workflow.

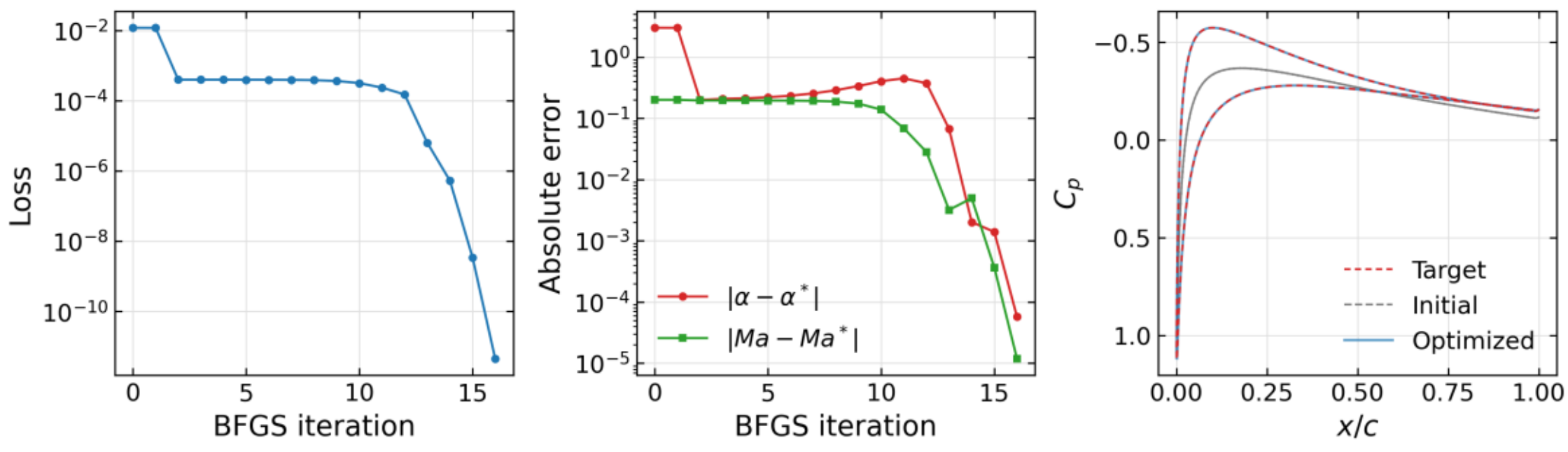

Figure 5. Freestream-parameter inversion using verified discrete-adjoint gradients.

## 4 Conclusions

This work revisits the practical value of explicit derivative kernels with agent-assisted programming. Unlike general-purpose automatic differentiation, explicit derivative kernels do not aim to provide black-box differentiation for arbitrary computational graphs. Instead, they target structured scientific-computing problems in which the required coordinate derivatives, residual Jacobian actions, and transpose-Jacobian actions are known in advance. In this setting, the role of agent assistance is to reduce the cost of generating, revising, and maintaining structured derivative code, rather than to replace mathematical derivation or numerical verification. The reliability of the generated kernels is established through independent consistency tests, including comparisons with automatic differentiation, Taylor-remainder tests, transpose inner-product checks, and discrete-adjoint gradient verification.

Two representative realizations were examined. For physics-informed neural networks, explicit differentiation (ED) reproduced the residual vectors and parameter gradients obtained by automatic differentiation (AD) to the corresponding floating-point accuracy for the Burgers, Kuramoto-Sivashinsky, and two-dimensional Navier-Stokes equations. Replacing coordinate AD inside the residual provided problem-dependent acceleration of complete 1000-step Adam training, often reaching 2-4× speedups in the more derivative-intensive cases, and reduced peak GPU memory in most tested settings. For the finite-volume CFD realization, the generated tangent-action and transpose-action kernels passed staged Taylor-remainder and inner-product consistency checks from local fluxes to global residual actions, boundary contributions, objective derivatives, and freestream-parameter derivatives. The same explicit-kernel strategy was then embedded into a GPU-resident discrete-adjoint workflow, where the verified derivative kernels supported Mach-number and angle-of-attack inversion in a conventional finite-volume solver. This result demonstrates that the proposed strategy can extend from neural-network coordinate derivatives to discretized PDE adjoints.

Overall, the results suggest that explicit derivative kernels should not be viewed only as costly and fragile hand-crafted implementations. When combined with agent-assisted implementation and stringent numerical verification, they can serve as a structured complement to automatic differentiation, providing a verifiable, structure-aware, and computationally efficient route for derivative-intensive scientific-computing tasks. A limitation of this approach is that the resulting derivative kernels are specialized to the prescribed network architecture, activation function, residual form, and discretization template. Changes in these components require the corresponding propagation rules or local tangent/transpose templates to be regenerated and reverified. Thus, the method does not provide black-box differentiation for arbitrary models or solvers; rather, it reduces the cost of constructing and maintaining verified problem-specific derivative kernels.

## Data Availability Statement

The benchmark data, verification scripts, and representative derivative-kernel implementations are available in the public GitHub repository: https://github.com/Cao-WenBo/ExplicitDifferentiation.

## Conflict of Interest Statement

The authors have no conflicts to disclose.

## Acknowledgments

We would like to acknowledge the support of the National Natural Science Foundation of China (Grant No.92152301).